\documentclass[useAMS,usenatbib,usegraphicx]{mn2e}
\usepackage{epsf}

\title[Dark and stellar matter in strong lensing galaxies ]{
Dark and stellar matter in strong lensing galaxies 
from a joint lensing and stellar dynamics\thanks{
Based on data collected at Subaru Telescope, which is operated by 
the National Astronomical Observatory of Japan.}}

\author[T. Hamana et al.]
{Takashi Hamana$^{1}$, Youichi Ohyama$^{2,1}$, Masashi Chiba$^3$,
Nobunari Kashikawa$^{1}$\\
$^1$ National Astronomical Observatory of Japan, 
Mitaka, Tokyo 181-8588, Japan\\
$^2$ Institute of Space and Astronautical Science, 
Japan Aerospace Exploration Agency, Sagamihara, Kanagawa
229-8510, Japan\\
$^3$ Astronomical Institute, Tohoku University,
Aoba-ku, Sendai 980-8578, Japan}
\date{Accepted ******; Received ******; in original form 2005 July 4}
\pagerange{\pageref{firstpage}--\pageref{lastpage}} 
\pubyear{2005} 
\volume{000}

\begin{document}

\label{firstpage}
\maketitle

%%%%%%%%%%%%%%%% Abstract
\begin{abstract}
We present Subaru spectroscopy of the early-type lensing galaxies in the lens 
systems HST~14113$+$5211 (at the redshift $z_L=0.464$) and B~2045$+$265
($z_L=0.868$), being aimed at measuring the velocity dispersion
of the lensing galaxies as an important component to their mass distributions 
and internal dynamics. 
For HST~14113$+$5211 we have obtained $174 \pm 20$ km~s$^{-1}$ (1~$\sigma$)
inside an aperture of 0.\arcsec9, and for B~2045$+$265 we have
obtained $213\pm 23$ km s$^{-1}$ inside an aperture of 1.\arcsec2.
To extract the significance of these information on the mass distributions and
stellar dynamics of the lensing galaxies, we construct two alternative models
for both the mass and velocity distributions.
It is found that the mass-to-light ratios as derived from a joint lensing 
and stellar dynamics analysis are virtually consistent with those 
obtained from the use of the Fundamental Plane. 
Also, for HST~14113$+$5211, we find that the total mass distribution is well
reproduced by a power-law density profile with an index of 2, thereby
suggesting that a singular isothermal model is a good fit to the lensing 
galaxy.
In contrast, the corresponding slope for B~2045$+$265 is flatter than 
isothermal,
suggesting that additional contributions to lensing mass from surrounding
structures (possibly the group of galaxies) provide the observed angular
separations of multiple images.
\end{abstract}

%%%%% keywords

%%%%%%%%%%%%%%% sec.1 Introduction
\section{Introduction}

Massive early type galaxies are predicted to form at the highest 
density peaks in the early universe and to evolve through subsequent 
mass assemblies.
Measuring their dynamical structure and stellar content over the cosmic 
time provides important constraints not only on their formation history 
itself but also on models of structure formation scenario and cosmic 
star formation history.

Studying the internal structure of early type galaxies is, however, difficult
because of lack of dynamical tracers at large radii such like H$_{\rm I}$ gas 
in spirals, and of degeneracy between kinematic properties of dynamical
tracers and mass distributions: 
The stellar dynamics suffers from the mass-anisotropy problem, and 
the gravitational lensing does from the mass-density profile problem
as well.

Those degeneracies can be broken by combining those two probes, 
since they nicely complement each other
(Treu \& Koopmans 2002; 2004 (hereafter TK04); Koopmans \& Treu 2003).
A combined analysis thus places important limits on
the distributions of luminous and dark matter in a lensing
early type galaxy.

We select two lens systems, HST14113$+$5211 ($z_L=0.464$; Fischer, 
Schade \& Barrientos 1998; Lubin et al. 2000) 
and B~2045$+$265 ($z_L=0.868$;  Fassnacht et al. 1999), 
where the lensing galaxies are an early-type galaxy. 
We conducted the direct measurement of velocity dispersion of the lensing 
galaxies of those systems using the Faint Object Camera and Spectrograph
(FOCAS, Kashikawa et al. 2002) mounted on the Subaru Telescope.
We perform a joint gravitational lensing and stellar dynamics analysis
to explore the luminous and dark matter distribution of lensing galaxies
focusing on the slope of (dark and total matter) density profile and
mass-to-light ratio.
Also, we utilize the fundamental plane relation of early type galaxies 
for deriving an alternative estimate of the mass-to-light ratio, 
which, after combined with the joint analysis, provides a useful 
constraint on the mass distribution of the lensing galaxy.

The paper is organized as follows.  
In \S 2, we show the observations and data reduction.
In \S 3, we derive the mass-to-light ratio of the lensing galaxies from
the fundamental plane relation of early type galaxies.
In \S 4, models for gravitational lensing and stellar dynamics are
presented. \S 5 is devoted to the results of our model analysis, and
discussion and concluding remarks are drawn in \S 6.

Throughout this paper, we adopt $\Omega_0=0.3$, $\lambda_0=0.7$, and
$h=H_0/100$ km~s$^{-1}$~Mpc$^{-1}$ $=0.65$ for the relevant estimations.

%%%%%%%% Sec.2 %%%
\section{VELOCITY DISPERSION MEASUREMENT}

%%%%% Sec 2-1
\subsection{Observations}

Spectroscopic observations were made with the Subaru 8.2m telescope 
(Iye et al. 2004).
The FOCAS spectrograph (Kashikawa et al. 2002) was configured with a 
0\arcsec.4-width slit, a 300 grooves mm$^{-1}$ grism which gives 
1.40\AA~per pixel, and a Y47 
order-cut filter to obtain optical spectra of the lensing galaxies of 
gravitational lens systems HST~14113$+$5211 and B~2045$+$265.
The on-chip binning was set to 3 (along spatial direction to give 
0\arcsec.3 per pixel) by 1 (along wavelength direction).

For HST~14113$+$5211, observation was made on 2002 June 13 and 14 (UT).
The slit was placed along the major axis of the lensing galaxy, 
whose position angles (PAs) was PA$=38.1^\circ$.
We obtained ten 1800 seconds exposures, and the total exposure 
time was 5 hours.
For B~2045$+$265, observation was made on 2002 June 13 (UT).
The slit was placed along the major axis of the lensing galaxy
(PA$=120.1^\circ$).
We obtained seven 1800 seconds exposures, and the total exposure 
time was 3.5 hours.
The seeing conditions were around 0.4\arcsec~-- 0.8\arcsec~during the 
two observing nights.

%%%%% Sec 2-2
\subsection{Data reduction}

Basic data reductions were made following Ohyama et al. (2002).
In the followings, we focus on the spectra of the lensing galaxies.
Wavelength accuracy and resolution of the galaxy spectra were measured, 
within the wavelength region for the Fourier cross-correlation analyses 
(see section 2.4), by 
means of Gaussian fittings of several narrowest (unblended) sky OH lines.
Here all measured values are shown in the rest frame of each galaxy, 
since all the following velocity dispersion measurements were made in 
their rest frames.
For the spectrum of HST~14113$+$5211, 
we found that wavelength resolution is 4.5\AA~FWHM at 
$4000-4450$\AA~(the blue fitting region) and 3.2\AA~at 
$4780-5170$\AA~(the red fitting region. See below for details 
of the spectrum fitting regions). The wavelength accuracy is found to be 
typically $\simeq 0.09$\AA~in RMS over $3800-6500$\AA, and it gets 
slightly worse to $\simeq 0.28$\AA~at the bluest wavelength 
($\sim 4000$\AA) for the cross-correlation analyses.
For B~2045$+$265, wavelength resolution is 2.7\AA~FWHM 
at $4100-4500$\AA, and wavelength accuracy is typically 
$\simeq 0.07$\AA~in RMS over $3800-4500$\AA.
Aperture size to extract an 1-dimensional galaxy spectrum are 
0\arcsec.9 (3-binned pixel, corresponding to 8.3 kpc at 
$z=0.464$ for the adopted cosmology) and 1\arcsec.2 
(4-binned pixel, 19 kpc at $z=0.868$) 
for HST~14113$+$5211 and B~2045$+$265, respectively, to obtain 
spectra with maximum signal-to-noise quality.

%%%%% Sec 2-3
\subsection{Basic properties of the spectra}

%%%%%%%% Figure 1
\begin{figure}
\includegraphics[width=60mm,angle=90]{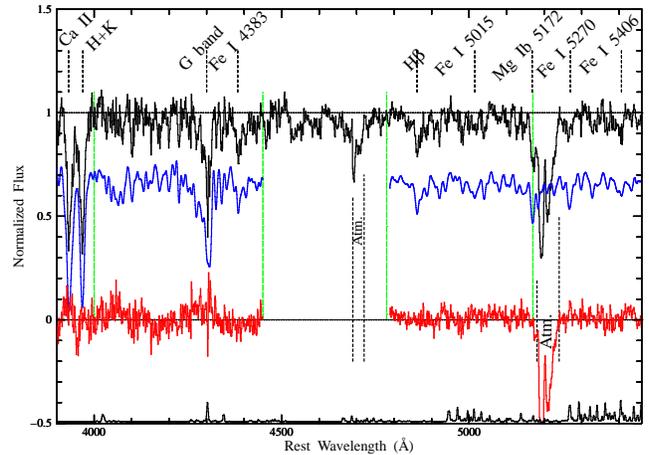}
\caption{Continuum-normalized spectrum of HST~14113$+$5211 ($black$) 
in the rest frame of the galaxy.
The continuum-normalized spectrum of a template star, HD 126778, 
after Gaussian-broadened to match the galaxy spectrum 
($\sigma = 174$ km s$^{-1}$), is shown ($blue$) with a vertical 
offset of $-0.3$ for clarity of the figure.
Residual spectrum (galaxy $-$ Gaussian-broadened 
template) ($red$) and a sky spectrum ($black$) are also shown.
Here the sky spectrum is shown, with a vertical offset of $-0.5$, 
to identify wavelength at which strong sky emission could damage 
the observed galaxy spectrum, and is flux-scaled by an arbitrary number.
Atmospheric extinction features are marked with vertical black broken lines.
Two wavelength regions used in the analyses of the velocity dispersion 
(blue and red fitting regions) are marked by two vertical $green$ lines 
(see a main text for details of the wavelength fitting regions).
Prominent features seen in the galaxy spectrum are identified as 
shown in upper portion of the figure.}
\label{fig:spectrum-hst}
\end{figure}

%%%%%%%% Figure 2
\begin{figure}
\includegraphics[width=60mm,angle=90]{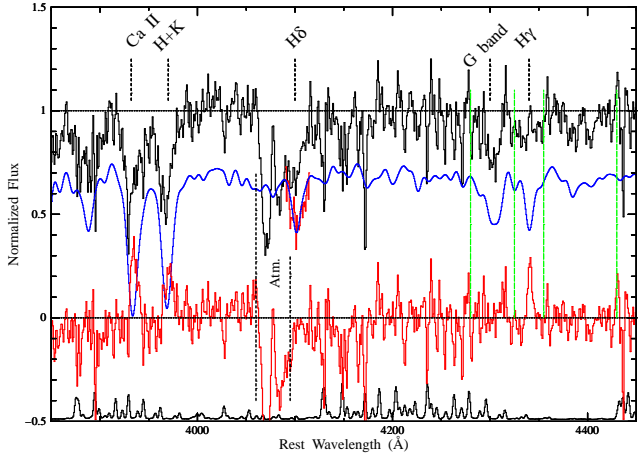}
\caption{Same as figure \ref{fig:spectrum-hst}, but for B~2045$+$265.
The sky spectrum is flux-scaled by a same amount as used for figure 
\ref{fig:spectrum-hst}.
Expected contribution of the galaxy absorption over the atmospheric 
A band absorption near 4100\AA~is shown in red, with a vertical 
offset of $-0.3$, overlaying the Gaussian-broadened 
($\sigma = 213$ km s$^{-1}$) template spectrum of HD 136202.
See a main text for details of this feature.}
\label{fig:spectrum-b}
\end{figure}

Reduced rest-frame spectra, after continuum normalization, are 
shown in Figures \ref{fig:spectrum-hst} and \ref{fig:spectrum-b}.
For HST~14113$+$5211, deep Ca II H and K absorption lines and 
prominent G band feature are evidently seen in the spectrum, 
suggesting that major contribution in the observed wavelength 
region could be attributed to the late type giant stars (late G 
giant -- early K giant stars).
The redshift of the lensing galaxy is measured to be $z=0.4644\pm 0.0002$ from 
these prominent features, and is consistent with previous 
measurement (Lubin et al. 2000).
In the redder part of the spectrum, several more features, 
including H$\beta$ and some Fe absorptions (Fe I 5270\AA~and 
5406\AA) are seen as well as Mg Ib 5172\AA~which is partly 
detected at the blue edge of the atmospheric absorption 
feature (the A band).

For B~2045$+$265, although signal-to-noise ratio of the observed spectrum 
is worse than that of HST~14113$+$5211 due to both fainter apparent 
brightness and shorter exposure time achieved for the galaxy, 
we have clearly detected Ca II H and K absorption lines and G 
band feature (Fig \ref{fig:spectrum-b}).
We note, however, that these features are less prominent than 
those of HST~14113$+$5211.
We also note that B~2045$+$265 shows another possible prominent 
absorption feature at just red edge of the A band feature.
We checked this possibility by comparing the A band features of 
both galaxies in the observed frame (at 7660\AA), and confirmed 
that B~2045$+$265 shows an excess absorption over expected A-band feature.
This feature, at 4100\AA~in the rest frame, can be identified as 
H$\delta$ absorption, although another Balmer absorption line, 
H$\gamma$ near the G band, was not detected.
Since H$\gamma$ emission is $\sim 1.8$ times brighter than 
H$\delta$ emission under the ``case B'' photoionization, 
which is typical for star-forming regions, difference in the 
observed properties of Balmer lines can be naturally understood.
Note that more higher-order Balmer absorptions were identified 
by Fassnacht et al. (1999).
Prominent [OII] emission is also detected, suggesting a star-forming 
activity in this galaxy.
Fassnacht et al. (1999) classified the spectrum as the Sa type,
and are best represented by giant stars of late 
F type.
The redshift of the lensing galaxy is measured to be $z=0.8682\pm 0.0001$ 
by [OII] emission, which is consistent with that measured by 
prominent absorption lines, although it is slightly larger than the value 
measured by Fassnacht et al. (1999) ($z=0.8673\pm 0.0005$).

%%%%% Sec 2-4
\subsection{Fourier cross-correlation analysis}

We basically followed the procedure of Ohyama et al. (2002), and 
the Fourier cross-correlation method
(Tonry \& Davis 1979) was 
used to measure the line-of-sight velocity dispersion of the 
lensing galaxies with the FXCOR task implemented in IRAF\footnote{
IRAF is distributed by the National Optical Astronomy Observatories,
which are operated by the Association of Universities for Research
in Astronomy, Inc., under cooperative agreement with the National
Science Foundation.} (see also Falco et al. 1997).
Here we outline our procedure applied for both galaxies.
Firstly, velocity-resolution matched stellar spectra were 
created from the ``C\'oude 
feed spectral library'' spectra (Leitherer et al. 1996), whose 
wavelength resolutions are 1.8\AA~FWHM for both red and 
blue spectra in the library, to match the resolution of the observed 
galaxy spectra, and were used as ``templates'' which 
are to be fitted with the observed galaxy spectra.
Secondly, calibration curves, which relate the velocity dispersion 
of Gaussian broadening function applied to the templates to the width 
of the cross-correlation functions (CCF) peak, were created.
Finally, curves were used to find the velocity dispersion of 
the galaxy from the CCF peak width calculated between the 
observed galaxy spectra and the templates.
FXCOR parameters were kept unchanged from that used by Ohyama et al. 
(2002) except for template stars and 
fitting wavelength regions to match the spectral properties 
of the program galaxies in this work.

%%%%%%%%%% Table 1
\begin{table}
\caption{Properties of template stars: 
(1) star name, 
(2) spectral type,
(3) effective temperature (K),
(4) surface gravity in log
and (5) metal abundance in units of solar metallicity.}
\label{table:stars}
\begin{tabular}{lcccc}
\hline
Star$^{(1)}$ & Type$^{(2)}$ & $T_{\rm eff}$$^{(3)}$ & log $g$$^{(4)}$ 
& [Fe/H]$^{(5)}$ \\
\hline
\multicolumn{5}{c}{for HST~14113$+$5211} \\
\hline
HD 206453 & G8III & 5092 & 2.14 & $-0.42$ \\
HD 1918   & G9III & 4865 & 2.01 & $-0.47$ \\
HD 126778 & K0III & 4847 & 2.34 & $-0.53$ \\
\hline
\multicolumn{5}{c}{for B~2045$+$265} \\
\hline
BD+11~2998 & F8III & 5425 & 2.30 & $-1.17$ \\
HD 136202  & F8III-IV & 6030 & 3.89 & $-0.07$ \\
\hline
\end{tabular}
\end{table}

%%%%% Sec 2-4-1
\subsubsection{HST~14113$+$5211}

%%%%%%%%% Table 2
\begin{table*}
\caption{Results of Fourier cross-correlation analyses for HST~14113$+$5211.
(1) The blue and red fitting regions are $4000-4450$\AA, and 
$4780-5170$\AA~respectively. 
(2) $R$ denotes Tonry-Davis $R$ value.}
\label{table:FCC-HST}
\begin{tabular}{cccccc}
\hline
 & & \multicolumn{2}{c}{blue$^{(1)}$} & \multicolumn{2}{c}{red$^{(1)}$} \\
Star & CCF fitting width & $\sigma$ (km s$^{-1})$ & $R$$^{(2)}$ 
& $\sigma$ (km s$^{-1})$ & $R$$^{(2)}$ \\
\hline
HD 206453 & 30 & 173 & 16.6 & 162 & 13.7 \\
          & 35 & 172 & 16.6 & 159 & 13.6 \\
HD 1918   & 30 & 175 & 15.5 & 179 & 14.0  \\
          & 35 & 172 & 16.6 & 170 & 14.1\\
HD 126778 & 30 & 191 & 17.8 & 170 & 14.7 \\
          & 35 & 190 & 18.0 & 168 & 14.8 \\
\hline
\multicolumn{6}{l}{$R$-weighted average $\sigma$:} \\
\multicolumn{1}{r}{for each fitting region} & 
& 179$\pm$9 km s$^{-1}$ & & 168$\pm$7 km s$^{-1}$ & \\
\multicolumn{1}{r}{for both fitting regions} & & 
& 174$\pm$10 km s$^{-1}$ & & \\
\hline
\end{tabular}
\end{table*}

For HST~14113$+$5211, we choose two separate wavelength regions 
($4000-4450$\AA~and $4780-5170$\AA, hereafter blue and 
red fitting regions, respectively) to be used in the FXCOR, 
which include most of the prominent absorption features, 
such as G band, Fe I, and H$\beta$, while avoiding atmospheric 
absorption features (A and B bands).
Fe I features at $5250-5410$\AA~are not included in the 
analyses because they are detected at redder part of the A 
band and available wavelength region around them is not wide enough.
The Ca II H and K features were also excluded from the fitting 
regions, since the calculation with these features might give 
problematic result because of their intrinsically wider line 
width (Tonry 1998).
We choose late type giant stars with sub-solar metallicity as 
spectral templates, since we found during the course of the 
spectrum fitting that stars with nearly solar metallicity 
show deeper Fe I absorptions than the observed ones.
We choose three such stars, HD 206453, HD 1918, and HD 126778 
(see Table \ref{table:stars} for their detailed properties)\footnote{
Metallicity of the selected template stars ([Fe/H]$=-0.4$ -- $-0.5$) 
is by no means the best estimated value for the galaxy.
They are arbitrarily chosen from a small subsample of low-metal stars 
in the spectral library.}.
For each star, blue ($3820-4500$\AA) and red ($4780-5450$\AA) 
spectra are available in the spectral library, and each of them was 
used separately in the FXCOR for fitting blue ($4000-4450$\AA) 
and red ($4780-5170$\AA) fitting regions, respectively.
Red and blue template spectra were processed separately in both the continuum 
normalization and velocity resolution matching procedures, 
and then each template was used separately in the FXCOR to 
make two calibration curves for a set of one template star and the galaxy.

We made FXCOR runs twelve times for each 
template (three stars), fitting region (two regions per star), and 
CCF peak width measurement lag (two lags, 30 and 35, per set of 
template and fitting region; see Ohyama et al. 2002 for details
of the parameter) to measure velocity dispersion of HST~14113$+$5211.
We found that all cross-correlations between the galaxy and templates 
gave good 
Tonry-Davis $R$ values ($R=14-18$; Tonry and Davis 1979), suggesting 
that templates adopted for HST~14113$+$5211 could represent the 
galaxy spectrum well.
We found that error of the galaxy velocity dispersion, originating 
from the error in measuring the CCF peak widths, was typically 
$\simeq 10$ km s$^{-1}$ ($1\sigma$).
Averaging all the FXCOR results with a weight of $R$s, we obtained 
$\sigma = 179\pm 9$ km s$^{-1}$ and $168\pm 7$ km s$^{-1}$ in blue 
and red fitting regions, respectively.
Here, errors represent $1\sigma$ scatters of the velocity dispersion 
values among all FXCOR runs.
Another source of error, the velocity resolution matching error, 
estimated from the error in measuring the spectral resolution of 
the galaxy spectrum, could amount to $\simeq 14$ km s$^{-1}$ and 
$\simeq 7$ km s$^{-1}$ of galaxy velocity dispersion error 
($1\sigma$) for blue and red fitting regions, respectively.
Considering all sources of uncertainties (errors in velocity-resolution 
matching, CCF peak width measurement, and scattering of the velocity 
dispersion values among FXCOR runs), the overall $1\sigma$ 
uncertainties are estimated to be 19 ($\sqrt{14^2+10^2+9^2}$) 
and 13 ($\sqrt{7^2+10^2+7^2}$) km s$^{-1}$ for 
blue and red fitting regions, respectively.
Note that results measured in two fitting regions are 
consistent to each other within expected errors.
Therefore, averaging all twelve FXCOR results made in blue and 
red fitting regions, we adopt $\sigma = 174\pm 20$ km s$^{-1}$ 
as the best estimated line-of-sight velocity dispersion of HST~14113$+$5211.
Here, an error is estimated, based on velocity matching error of 
14 km s$^{-1}$ (larger value out of two fitting regions), 
CCF fitting error of 10 km s$^{-1}$, and scattering among 
all FXCOR results (10 km s$^{-1}$; see Table \ref{table:FCC-HST}), 
to be 20 ($\sqrt{14^2+10^2+10^2}$) km s$^{-1}$.

Fig \ref{fig:spectrum-hst} compares the spectrum of 
HST~14113$+$5211 with the best 
template, HD 126778, giving the best $R$ values for both blue and 
red fitting regions, after Gaussian-broadened with velocity 
dispersion of $\sigma = 174$ km s$^{-1}$.
We also show residual spectrum (galaxy $-$ Gaussian-broadened 
template) along with the sky spectrum.
All the spectra are shown over wider wavelength region, covering 
other spectral features not used for the FXCOR analyses, to show 
the overall quality of the spectral fitting.
One may find that the broadened template gives rather good 
representation of the observed galaxy spectrum over entire 
wavelength region, except for a large-scale residual pattern 
in the blue continuum.
The residual feature likely comes from errors in fitting the 
continuum spectra of both the galaxy and the templates, used 
for continuum normalization.
Since the continuum fittings were made over the wavelength regions 
where spectra show rather large-amplitude change in shape, an 
artificial large-scale spectral variation could be easily introduced 
into the continuum-normalized spectra.
Note, however, that this error does not affect our cross-correlation 
results, since such a low-frequency variation in spectral shape is 
filtered out in the FXCOR before calculating the CCF (see Ohyama 
et al. 2002 for more details on the FXCOR parameters).
In concluding, after considering the fitting quality as discussed 
above, we finally adopt $\sigma = 174\pm 20$ km s$^{-1}$ as the best 
estimated velocity dispersion of HST~14113$+$5211 in the following 
lens analysis.

%%%%% Sec 2-4-2
\subsubsection{B~2045$+$265}

%%%%%%%%%%%% Table 3
\begin{table*}
\caption{Same as Table \ref{table:FCC-HST} but for B~2045$+$265.
Fitting region is $4280-4325$\AA~plus 
$4355-4430$\AA.}
\label{table:FCC-B}
\begin{tabular}{cccc}
\hline
Star & CCF fitting width & $\sigma$ (km s$^{-1})$ & $R$ \\
\hline
BD+11~2998 & 30 & 219 & 30.9 \\
           & 35 & 234 & 30.5 \\
HD 136202  & 30 & 210 & 39.9 \\
           & 35 & 213 & 39.7 \\
\hline
$R$-weighted average $\sigma$: & & 213$\pm$11 km s$^{-1}$ & \\
\hline
\end{tabular}
\end{table*}

For B~2045$+$265, we followed the same procedure used for HST~14113$+$5211.
However, because of relatively poor quality of B~2045$+$265 spectrum, 
only a narrow wavelength region, comprising two smaller wavelength 
regions of $4280-4325$\AA~and $4355-4430$\AA, was chosen for 
the FXCOR analyses.
The spectral regions were carefully chosen to include G band while 
avoiding both the strong sky emission regions and H$\gamma$, and 
these two smaller regions were fitted at a same time in FXCOR.
Here, the H$\gamma$ was excluded in the fitting since H$\gamma$ 
emission may overlay underlying H$\gamma$ absorption.
We choose two late F giant stars, BD+11~2998 and HD 136202, as 
spectral templates for the galaxy from the same spectral library 
(see Table \ref{table:stars} for their detailed properties)\footnote{
Again, as in the case of the analyses of HST~14113$+$5211, metallicity 
of the selected template stars is by no means the best estimated value 
for the galaxy.}.
We ran FXCOR four times (for one fitting region, two template 
stars, and with two CCF peak width measurement lags), and found 
that $R$ values are rather large, $30-40$, for all FXCOR runs.
The larger $R$ values probably result from the fact that we have 
only one prominent spectral feature (G band) in a narrow 
wavelength region for fitting.
The best estimated velocity dispersion is $\sigma = 213$ km s$^{-1}$, 
with an $1\sigma$ error of $\pm 11$ km s$^{-1}$ originating from scattering of 
the $\sigma$ values among all FXCOR runs (see Table \ref{table:FCC-B}).
The velocity resolution-matching error could amount to $\simeq 5$ 
km s$^{-1}$ of galaxy velocity dispersion error ($1\sigma$), 
which is smaller than in the case of HST~14113$+$5211 due to 
higher redshift of B~2045$+$265.
We found that error of the galaxy velocity dispersion, originating 
from the error in measuring the CCF peak widths, was typically 
$\simeq 19$ km s$^{-1}$ ($1\sigma$), which is larger than the value 
for HST~14113$+$5211 because of relatively poor CCF determination 
resulting from lower quality spectrum and narrower wavelength fitting 
regions available for B~2045$+$265.
Considering all sources of uncertainties (errors in velocity-resolution 
matching, CCF peak width measurement, and scattering of the velocity 
dispersion values among FXCOR runs), the overall $1\sigma$ 
uncertainty is estimated to be 23 ($\sqrt{5^2+19^2+11^2}$) km s$^{-1}$.

Fig \ref{fig:spectrum-b} compares the observed galaxy spectrum with the 
Gaussian-broadened template spectrum of the best template, HD 136202, 
after Gaussian-broadened with the velocity dispersion of $\sigma =213$ 
km s$^{-1}$, as in the same way for HST~14113$+$5211.
One may find that the broadened template gives rather good representation 
of the galaxy spectrum over entire wavelength region, including 
H$\delta$, except for H$\gamma$ and Ca II H and K features.
We expect that the residual emission of H$\gamma$ represents the real 
emission, overlaying on the absorption line, as expected from another 
emission line of [OII].
On the other hand, it seems difficult to attribute the 
shallower-than-observed Ca II absorptions in the broadened template spectrum
to overlaying emissions.
Although putative H$\epsilon$ emission, whose wavelength is almost 
coincident with that of Ca II H absorption, might explain a part of 
the residual feature, there seems no emission lines which could be 
coincident with Ca II K feature neither in the spectra of the lens 
galaxy nor the lensed QSO.
Therefore, the most simple and plausible explanation for the shallower 
observed Ca II absorption is that the lensing galaxy contains not only 
stars of late F type but also of even earlier types.
Although neglecting the contribution of the earlier population in the 
FXCOR analyses should affect the velocity dispersion measurement, we 
expect that our result, made assuming a pure stellar population of 
late F type as spectral templates, is still reasonably reliable and is
useful for our analyses 
on the mass model and the stellar dynamics in the following sections.
This is because (1) stars of earlier types show much shallow G band 
feature, and they can not fit the observed spectrum alone, indicating that 
late F stars contribute most to the observed spectrum in flux, and (2) 
our measurement almost relies on the spectral fit of the G band feature 
and, hence, the velocity dispersion measured in our analyses lies larger 
weight on the later type stars, which contribute much more on the mass 
of the lensing galaxy than the case of stars of earlier types.
Therefore, we adopt $\sigma = 213\pm 23$ km s$^{-1}$ as the best 
estimated line-of-sight velocity dispersion of B~2045$+$265 in the 
following lens analysis.

%%%%%%%%%% Sec 3
\section{Fundamental plane}

Before going into a joint gravitational lensing and stellar dynamics,
let us examine the Fundamental Plane (FP) for early type galaxies, which 
provides us with an alternative insight into the value of the 
mass-to-light ratio $M_\ast/L_B$. 
The FP is defined by the effective radius
($R_e$ kpc), central velocity dispersion ($\sigma_c$ km~s$^{-1}$),
and mean surface brightness inside $R_e$ ($SB_e$ mag~arcsec$^{-2}$):
\begin{equation}
\log R_e = \alpha_{FP} \log \sigma_c + \beta_{FP} SB_e + \gamma_{FP} \ ,
\label{eq:FP}
\end{equation}
where $\alpha_{FP}=1.25$, $\beta_{FP}=0.32$, and
$\gamma_{FP} = -8.895 - \log h_{50}$ ($h_{50}=H_0/50$ km~s$^{-1}$~Mpc$^{-1}$)
in the B band (Bender et al. 1998).
The central velocity dispersion $\sigma_c$ is taken as the velocity dispersion
measured inside $R_e/8$ (TK04).

For the lensing galaxy of HST~14113$+$5211,
Fischer, Schade, \& Barrientos (1998, hereafter F98)
obtained the rest-frame central surface brightness of
the lensing galaxy as $\mu_{0,B} (AB) = 13.57$ mag~arcsec$^{-2}$. Then,
as the surface brightness at $R_e=0.\arcsec 61 \pm 0.\arcsec 03$ (F98)
is $\mu_{e,B} (AB) = \mu_{0,B} (AB) + 8.33$
and $B({\rm Vega}) = B(AB) + 0.12$, we obtain the surface brightness
at $R_e$ as $\mu_{e,B} = 22.02$ mag~arcsec$^{-2}$. The mean surface brightness
within $R_e$ is then estimated as $SB_e = 20.56$ mag~arcsec$^{-2}$ after
correction for Galactic extinction $E(B-V)=0.016$ mag
(Schlegel, Finkbeiner, \& Davis 1998) and an $R_V=3.1$ extinction curve.
Alternatively, van~de~Ven, van~Dokkum, \& Franx (2003, hereafter vdV03) derived
$\mu_{e,B} = 21.70$ mag~arcsec$^{-2}$ at the intermediate effective
radius $R_e$ of $0.\arcsec 48$ (Kochanek et al. 2000), giving
$SB_e = \mu_{e,B} - 1.393 \simeq 20.56$ mag~arcsec$^{-2}$.
As the small difference between these values results in only the difference
of 0.2 to 0.3 in the mass-to-light ratio obtained below, we adopt the
results of F98 in what follows.

For the lensing galaxy of B~2045$+$265, we adopt the values given in vdV03,
i.e., $R_e = 0.\arcsec 38^{+0.\arcsec 15}_{-0.\arcsec 11}$
and $SB_e = \mu_{e,B} - 1.393 \simeq 19.11 \pm 0.44$ mag~arcsec$^{-2}$.

To derive a fiducial correction factor from the measured $\sigma$
in a specific aperture to $\sigma_c$ in equation (\ref{eq:FP}), we utilize
a fiducial model for internal mass distribution and stellar dynamics
as explained in \S \ref{sec:models}. Briefly saying, we adopt a single mass
component represented by a power-law density profile with an index of
$\gamma'$ [see equation (5)] and the velocity dispersion of stars
represented by a constant anisotropy parameter $\beta$ [see equation (6)].
For HST~14113$+$5211, we obtain $\sigma_c = 1.15 \sigma$
based on the most likely parameters of $(\gamma',\beta)=(1.93,0.14)$
(see \S \ref{sec:results})
thereby yielding $\sigma_c = 200.7$ km~s$^{-1}$. 
For B~2045$+$265, we obtain $\sigma_c = 0.99 \sigma$
based on the most likely parameters of $(\gamma',\beta)=(1.66,0.10)$,
thereby yielding $\sigma_c = 210.3$ km~s$^{-1}$.

Inserting into equation (\ref{eq:FP}), we obtain
$\gamma_{FP} = -8.87$
for the lensing galaxy of HST~14113$+$5211 at $z_L=0.464$ and
$\gamma_{FP} = -8.52$ for the lensing galaxy of B~2045$+$265 at $z_L=0.87$.
Comparison with $\gamma_{FP}$ at $z=0$ allows us to derive the difference
in $\gamma_{FP}$, $\Delta \gamma_{FP}$, and then
$\Delta \log (M_\ast/L_B) \equiv -0.4 \Delta \gamma_{FP} / \beta_{FP}$
provided $\alpha_{FP}$ and $\beta_{FP}$ are constant.
We then obtain $\Delta \log (M_\ast/L_B) = -0.172$
for the lensing galaxy of HST~14113$+$5211
and $-0.612$ for the lensing galaxy of B~2045$+$265.
The mass-to-light ratio at $z>0$ is given as,
\begin{equation}
\log (M_\ast/L_B)_z = \log (M_\ast/L_B)_0 + \Delta \log (M/L_B)
\end{equation}
where $(M_\ast/L_B)_0 = 7.3 \pm 2.1 h_{65}$~$M_\odot/L_{B,\odot}$
(Gerhard et al. 2001). We obtain
$(M_\ast/L_B)_z = 4.9 \pm 1.4$ $M_\odot/L_{B,\odot}$
for the lensing galaxy of HST~14113$+$5211 and
$1.8 \pm 0.5$ $M_\odot/L_{B,\odot}$ for the lensing galaxy of B~2045$+$265.
As will be shown below, these values of $M_\ast/L_B$ for both systems
are in good agreement with those obtained from a joint lensing and 
dynamical analysis (see \S \ref{sec:results}).

%%%%%%%%% Sec.4 %%%
\section{MASS MODEL AND STELLAR DYNAMICS}
\label{sec:models}

%%%%%%%% Sec 4.1 %%%
\subsection{Lens models}

%%%%%%%% Figure 3
\begin{figure}
\begin{center}
\begin{minipage}{8.4cm}
\epsfxsize=8.4cm 
\epsffile{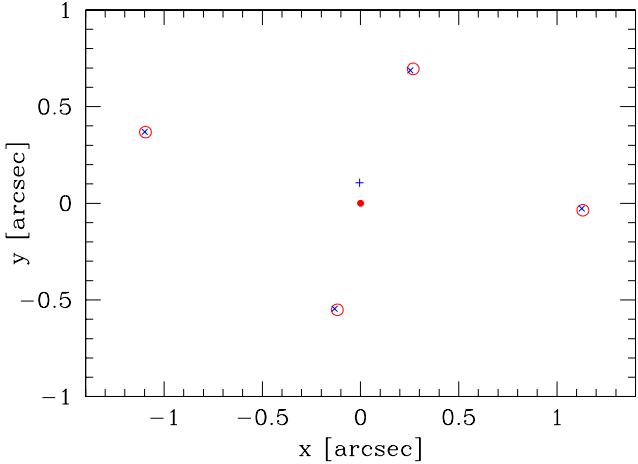}
\end{minipage}
\end{center}
\caption{Lens configuration of the HST~14113$+$5211 compared with SIE 
lens with an external shear model.
Multiple image positions and source position of the best-fitting model are
shown by crosses and plus, respectively.
Open circles show the observed image positions, where the radii of the circles
correspond to 1-$\sigma$ observational uncertainties.
The lensing galaxy position is shown by filled circle and is taken as
the coordinate origin. }
\label{fig:hst_lensconfig}
\end{figure}

%%%%%%%% Figure 4
\begin{figure}
\begin{center}
\begin{minipage}{8.4cm}
\epsfxsize=8.4cm 
\epsffile{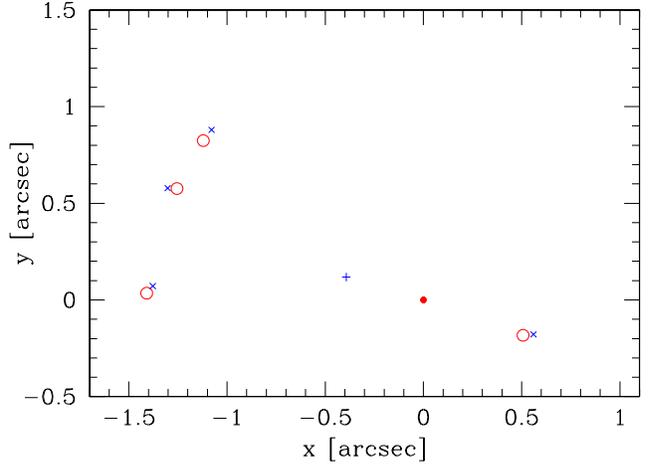}
\end{minipage}
\end{center}
\caption{Same as Fig.~\ref{fig:hst_lensconfig} but for B~2045$+$265.}
\label{fig:b2045_lensconfig}
\end{figure}

Following TK04, we model the lensing galaxy as a
singular isothermal ellipsoid (SIE: Kormann, Schneider \& Bartelmann 1994).
Note that the Einstein radius ($R_E$) and mass enclosed by the Einstein radius
($M_E$), both of which are the quantities required in the dynamical model
of the lensing galaxy in the following sections, are very insensitive to the
assumed mass model (Kochanek 1991; Koopmans \& Treu 2004).
We also allow for a constant external shear.
The observed position of a lensing galaxy is taken by a lens position
and we do not treat the lens position as a free parameter.
Therefore, our lens model has five parameters (we follow the notations
in TK04); the lensing strength 
($b_l= 4 \pi (\sigma_{\rm SIE} /c )^2 D_{ls}/D_s$, where $\sigma_{\rm SIE}$
is the one-dimensional velocity dispersion of SIE lens), axis ratio
($q_l$), position angle of the lens ($\theta_l$), strength of the external
shear ($\gamma_{\rm ext}$), and its orientation ($\theta_{\rm ext}$).
Also we treat the source position ($\beta_x$, $\beta_y$) as a free parameter.

%%%%%  Table 3
\begin{table}
\caption{Best-fitting model parameters and characteristic values of 
the singular isothermal ellipsoid with an external shear lens
\label{table:best-model}}
\begin{tabular}{lcc}
\hline
Parameter        & HST~14113$+$5211 & B~2045$+$265 \\
\hline
$b_l$ (arcsec)            & 0.84    & 1.11 \\
$q_l$                     & 0.68    & 0.68 \\
$\theta_l$ (deg)          & 38      & $-69$ \\
$\gamma_{\rm ext}$        & 0.26    & 0.05 \\
$\theta_{\rm ext}$ (deg)  & $-35$     & 18 \\
$\beta_x$ (arcsec)        & $-0.005$  & $-0.39$ \\
$\beta_y$ (arcsec)        &  0.11   & 0.12 \\
$\sigma_{\rm SIE}$ (km/s) & 202     & 397 \\
$R_E$ (kpc)               & 5.29    & 9.19 \\ 
$M(<R_E) (M_\odot)$ & $1.58\times 10^{11}$ & $1.06\times 10^{12}$ \\
\hline
\end{tabular}
\end{table}

We search for a set of model parameters that best reproduces the observed
lens configuration.
The best-fitting model parameters are summarized in Table 
\ref{table:best-model}.
Figs \ref{fig:hst_lensconfig} and \ref{fig:b2045_lensconfig} 
compare the observed image positions with the model images.
The usual $\chi^2$ values are 0.71 and 14 for HST14113$+$5211 and 
B~2045$+$265, 
respectively (for 1-$\sigma$ observational uncertainties
of 0.03 arcsec) and the number of degrees of freedom is $N_{dof}=1$
(8 constraints and 7 parameters).
As is evidently shown in Fig \ref{fig:hst_lensconfig}, 
for HST14113$+$5211, the model reproduces the 
lens configuration very well. 
The Einstein radius and mass enclosed by the Einstein radius are found to be
$R_E$=5.29kpc and $M(<R_E)=1.58\times 10^{11}M_\odot$ (for the adopted
cosmological parameters), respectively. 
On the other hand, for B~2045$+$265, although the model nicely reproduces 
the cusp-lensing configuration (Schneider Ehlers \& Falco 1992), the 
best-fitting model positions slightly deviate from the observed positions.
Since the enclosed mass is not very sensitive to a detail lens 
model but is primarily determined by the image separation, the best-fitting 
model may give a good estimate of the the enclosed mass.
The Einstein radius and mass enclosed are found to be
$R_E$=9.19kpc and  $M(<R_E)=1.06\times 10^{12}M_\odot$, respectively,
which we adopt in the following joint lensing and stellar dynamics analysis.

%%%%%%%%%% Sec 4-2
\subsection{Mass model and stellar dynamics}

Let the luminous-mass density and dark-matter density profiles be
$\rho_{lum}(r)$ and $\rho_{DM}(r)$, respectively, provided that these are
spherically symmetric. Two alternative mass models will be employed,
following TK04.

First, we consider a two-component mass model, where the luminous mass and
dark matter distribute differently:
\begin{eqnarray}
\rho_{lum}(r) &=& \frac{M_\ast r_\ast}{2\pi r (r + r_\ast)^3} \\
\rho_{DM}(r)  &=& \frac{\rho_{DM,0} r_b^3}{r^\gamma
                      (r^2 + r_b^2)^{(3-\gamma)/2}} \ ,
\end{eqnarray}
where $M_\ast$ is the total stellar mass and $r_\ast$ denotes the scale
length for the luminous matter. The profile $\rho_{lum}(r)$ corresponds to
a Hernquist model (Hernquist 1990), reproducing the $R^{1/4}$ surface 
brightness profile
with an effective radius of $R_e = 1.8153 r_\ast$. For $M_\ast$, we use
the $B$-band total luminosity of the lensing galaxy, $L_B$, based on the
mass-to-light ratio of the luminous matter, $M_\ast/L_B$, as a model parameter.
Our estimation of $L_B$ is given in Appendix, yielding $L_B/L_{B,\odot} =
2.6 \times 10^{10}$ and $1.3 \times 10^{11}$ ($h=0.65$) for
the lensing galaxies of HST~14113$+$521 and B~2045$+$265, respectively.
The dark-matter density profile, $\rho_{DM}(r)$,
is determined by the scale length ($r_b$),
density scale ($\rho_{DM,0}$), and inner slope ($\gamma$). The combination
of all of these parameters will be constrained by the results of the lens
fitting, $R_E$ and $M(<R_E)$, as obtained in \S 4.1.

Second, we consider a single component with power-law mass model:
\begin{equation}
\rho_{tot}(r) \propto r^{-\gamma'} \ ,
\end{equation}
where $\gamma'$ denotes an effective slope.

The velocity distribution of the stars is also assumed
to be spherically symmetric, such that the velocity dispersions in spherical
coordinates $(\sigma_r, \sigma_\theta, \sigma_\phi)$ satisfy
$\sigma_\theta = \sigma_\phi$ (e.g., Binney \& Tremaine 1987). We then use
the parameter $\beta(r) = 1 - \sigma_\theta^2 / \sigma_r^2$ describing
the degree of velocity anisotropy.
We employ the following two models for $\beta(r)$,
the Osipkov-Merritt model with a parameter $r_i$ (referred to as Model A) and
constant anisotropy model with a parameter $\beta$ (Model B).
\begin{equation}
\beta(r) = \left\{
 \begin{array}{ll}
\frac{r^2}{r^2 + r_i^2} , &\quad\mbox{for Model A} \\
const.(=\beta)          , &\quad\mbox{for Model B}
 \end{array}\right.
\end{equation}
Relevant parameters are $r_i$ and $\beta$.

%%%%%%%%%% Sec.5 %%%
\section{Results of a Joint Lensing and Dynamical Analysis}
\label{sec:results}

%%%%%%%%%% Figure 5
\begin{figure}
\begin{center}
\begin{minipage}{8.4cm}
\epsfxsize=8.4cm 
\epsffile{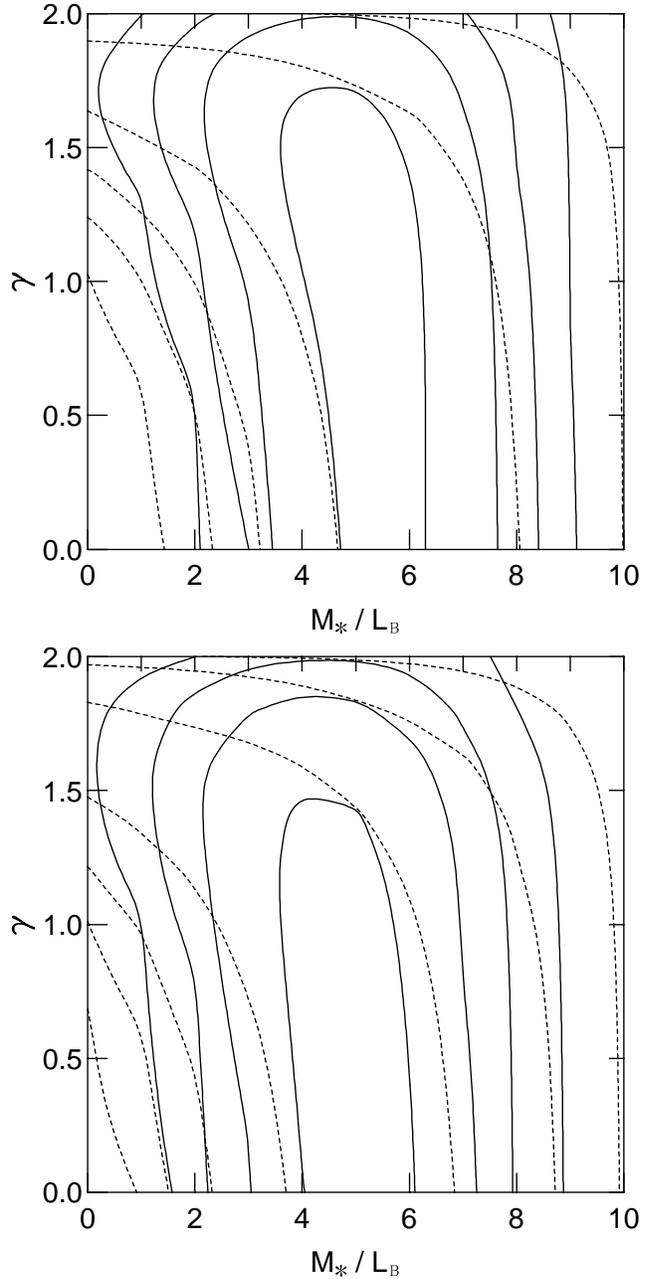}
\end{minipage}
\end{center}
\caption{
Likelihood contours for a joint lensing and dynamical analysis of
HST~14113$+$5211 based on the two-component mass model.
Contours show 68 \% (1$\sigma$), 95 \%, 99 \%, and 99.9 \%
confidence levels. Dashed lines are derived from the current joint analysis,
while solid lines combine the additional constraints from
the FP plane.
Lower panel: Model A with $r_i=R_e$ and $r_b=R_E$.
Upper panel: Model B with $\beta=0$ and $r_b=R_E$.}
\label{fig:likeli1}
\end{figure}
%%%%%%%%%%

%%%%%%%%%% Figure 6
\begin{figure}
\begin{center}
\begin{minipage}{8.4cm}
\epsfxsize=8.4cm 
\epsffile{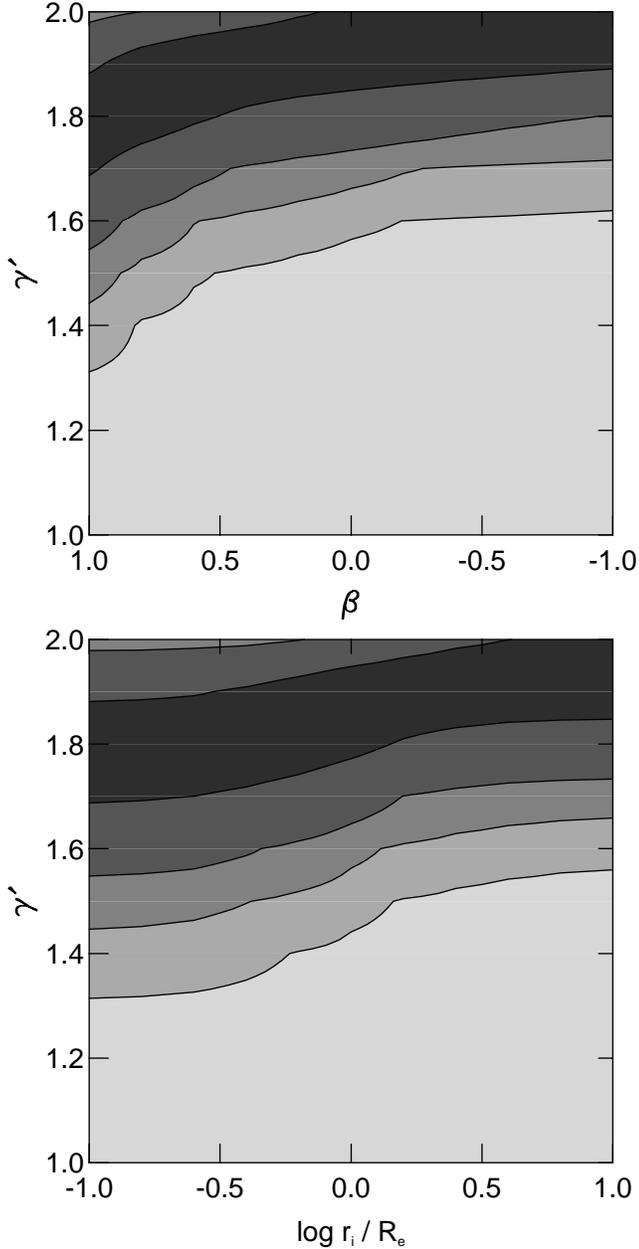}
\end{minipage}
\end{center}
\caption{
Likelihood contours for a joint lensing and dynamical analysis of
HST~14113$+$5211 based on the single component with power-law mass model.
Contours show 68 \% (1$\sigma$), 95 \%, 99 \%, and 99.9 \% confidence levels.
Lower and upper panels are for Model A and Model B, respectively.
\label{fig:likeli2}}
\end{figure}
%%%%%%%%%%

Based on the models given in the previous section, we search for the best
mass model for reproducing the observed velocity dispersion $\sigma$
using the joint lensing and dynamical analysis and
also combining with the FP constraints.
We especially focus on the best values of $M_\ast/L_B$ for a two-component
model, which are to be compared with those from the FP constraints.
Also, the inner slope of the dark-matter halos, $\gamma$, are to be compared
with the prediction of cosmological N-body simulations
[e.g., $\gamma = 1-1.5$ (Navarro, Frenk, \& White 1995; Moore et al. 1998; 
Fukushige \& Makino 2003 and references therein)]. 
For a single-component model, the derived
slope $\gamma'$ can be used to assess the isothermality of the lens density
profile as usually adopted in other lensing work.

%%%%% Sec 5-1
\subsection{HST~14113$+$5211}

First, we examine a two-component mass model. Figure \ref{fig:likeli1} shows
68 \% (1$\sigma$), 95 \%, 99 \%, and 99.9 \% confidence levels of
the likelihood contours in the plane of $\gamma$ and $M_\ast/L_B$
when we adopt the velocity dispersion of $\sigma=174 \pm 20 $ km~s$^{-1}$.
Dashed lines are derived from the joint lensing and dynamical analysis,
while solid lines combine the additional constraints from the FP plane.
For the lower panel (Model A) we set $r_i=R_e$ and $r_b=R_E$, and
for the upper panel (Model B) we set $\beta=0$ and $r_b=R_E$.
The likely $\gamma$ or $M_\ast/L_B$ using these parameter sets will be
utilized as a characteristic case in what follows, as the likelihood
contours are found to remain basically the same in other parameter
settings (TK04).

{}From the results based on the current joint analysis (dashed lines),
the likely mass-to-light ratio appears to be in the range of
$4 \la M_\ast/L_B \la 6$ $M_\odot/L_{B,\odot}$:
after marginalizing over $\gamma$, we obtain
the most likely values of $M_\ast/L_B$ ($M_\odot/L_{B,\odot}$), yielding
$4.4^{+2.0}_{-2.1}$ for Model A and $5.6^{+2.2}_{-2.2}$ for Model B.
As is evident, these values of $M_\ast/L_B$ are virtually
consistent with those obtained from the FP in \S 3
($4.9 \pm 1.4$ $M_\odot/L_{B,\odot}$).
We also calculate the most likely fraction of the dark matter projected inside
$R_E$ ($\simeq 1.4 R_e$), which is denoted as $f_{DM}$. Adopting the isotropic
velocity model as a representative case, we obtain
$f_{DM} = 0.47^{+0.21}_{-0.21}$, suggesting that about a half of
the total mass derived from the lens model constitutes dark matter.

Turn next to the result from the joint analysis combined with 
the FP constraints (plotted by the solid lines).
Since the above joint analysis gives $M_\ast/L_B$ being in a good agreement
with one from the FP, the additional constraint from the FP makes only a
minor change in the preferred $M_\ast/L_B$ value.
It is found that marginalized constraints on $M_\ast/L_B$ 
($M_\odot/L_{B,\odot}$) are $4.8^{+1.1}_{-1.2}$ for Model A and 
$5.1^{+1.2}_{-1.1}$ for Model B.
The most likely fraction of the dark matter projected inside $R_E$
(for the isotropic velocity case)
reads $f_{DM} = 0.52^{+0.10}_{-0.12}$.
The additional constraint from the FP improves the limit on $\gamma$.
We obtained, after marginalizing over $M_\ast/L_B$, a constraint on 
$\gamma$ to be $\gamma < 1.6$ ($\gamma < 1.8$) at 1-$\sigma$ 
for Model A (Model B).

Second, we examine a single-component mass model to set limits on the slope
$\gamma'$. Figure \ref{fig:likeli2} shows
68 \% (1$\sigma$), 95 \%, 99 \%, and 99.9 \% confidence levels of
the likelihood contours.
We set $r_b=3R_E$ in this diagram, but the limits on
$\gamma'$ are found to be little affected as long as $r_b \ga 3 R_E$.
We estimate the most likely values of $\gamma'$ for $r_i=R_e$ (Model A)
and $\beta=0$ (Model B), yielding $1.87^{+0.08}_{-0.09}$ and
$1.94^{+0.07}_{-0.09}$, respectively.
Thus, the total density profile is well approximated as
$\rho_{tot}(r) \propto r^{-2}$.

%%%%%%%%%% Figure 7
\begin{figure}
\begin{center}
\begin{minipage}{8.4cm}
\epsfxsize=8.4cm 
\epsffile{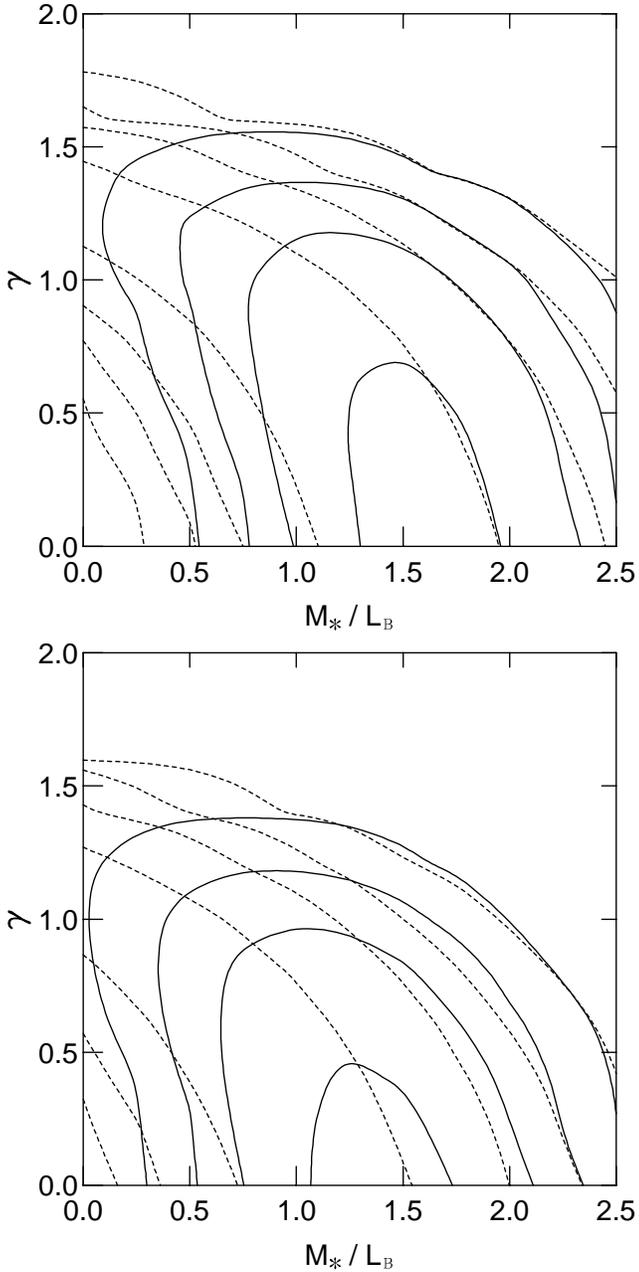}
\end{minipage}
\end{center}
\caption{
The same as Fig.\ref{fig:likeli1} but for B~2045$+$265.}
\label{fig:B2045_likeli1}
\end{figure}
%%%%%%%%%%

%%%%%%%%%% Figure 8
\begin{figure}
\begin{center}
\begin{minipage}{8.4cm}
\epsfxsize=8.4cm 
\epsffile{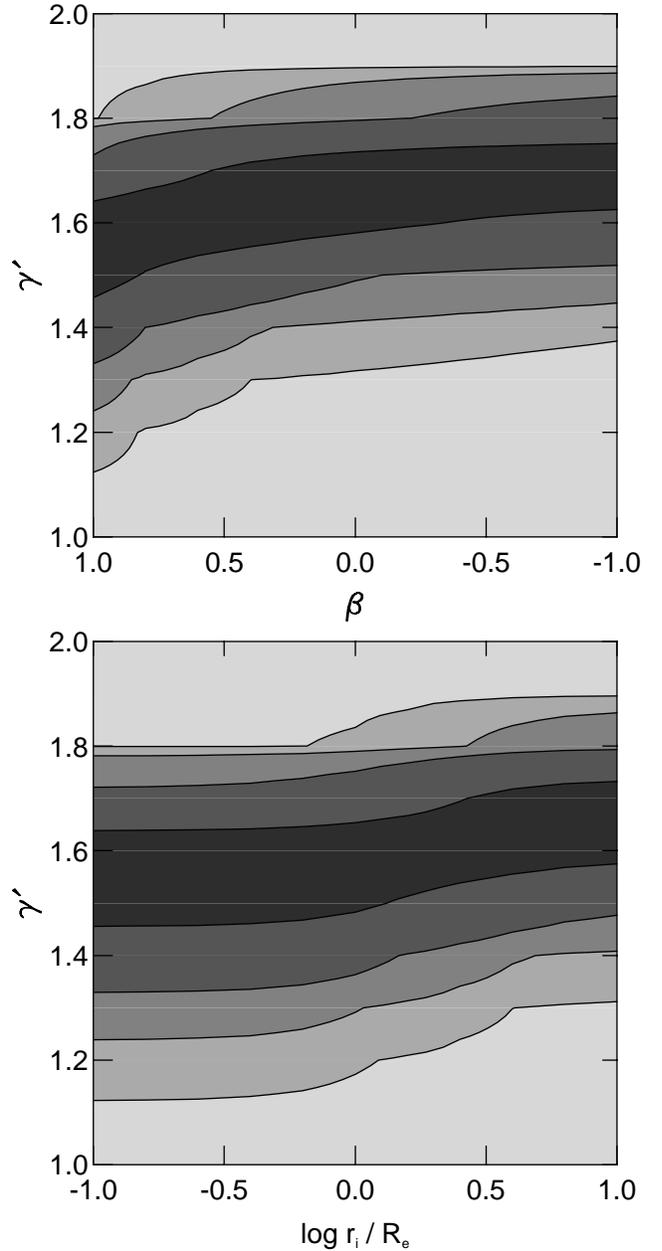}
\end{minipage}
\end{center}
\caption{
The same as Fig.\ref{fig:likeli2} but for B~2045$+$265.
\label{fig:B2045_likeli2}}
\end{figure}
%%%%%%%%%%

%%%%% Sec 5-2
\subsection{B~2045$+$265}

For the lensing galaxy of B~2045$+$265, we show the results in Figure
\ref{fig:B2045_likeli1} and \ref{fig:B2045_likeli2}, which are to be compared
with Figure \ref{fig:likeli1} and \ref{fig:likeli2}, respectively, obtained
for HST~14113$+$5211.

For a two-component mass model (Figure \ref{fig:B2045_likeli1}), 
the likely mass-to-light ratio appears to be in the range of
$0.7 \la M_\ast/L_B \la 2$ $M_\odot/L_{B,\odot}$:
after marginalizing over $\gamma$, we obtain
the most likely values of $M_\ast/L_B$ ($M_\odot/L_{B,\odot}$) for
Model A (Model B) as $M_\ast/L_B = 0.8^{+0.5}_{-0.7}$
($1.2^{+0.5}_{-0.7}$) without the FP constraints.
As is evident, these values of $M_\ast/L_B$ are also
consistent with those obtained from the FP in \S 3
($1.8 \pm 0.5$ $M_\odot/L_{B,\odot}$).
Combining the FP constraint and after marginalizing over $M_\ast/L_B$
we obtain $\gamma<0.5$ ($\gamma<0.8$) at $1~\sigma$ for Model A (Model B).
Thus the models prefer flatter dark matter inner slope than
the case of HST~14113$+$5211.

We also obtain the most likely fraction of the dark matter projected inside
$R_E$ ($\simeq 2.9 R_e$) for the isotropic velocity case,
yielding $f_{DM} = 0.89^{+0.06}_{-0.04}$ (without the FP
constraints) and $0.86^{+0.05}_{-0.03}$ (with the FP constraints).
Thus, in contrast to the case of
HST~14113$+$5211, the total mass inside $R_E$ derived from the lens model
is totally dominated by dark matter; some of this unseen mass
component may be provided by the group of galaxies which the lensing galaxy
belongs to.

For a single-component mass model (Figure \ref{fig:B2045_likeli2}),
we obtain the most likely values of $\gamma'$ for $r_i=R_e$ (Model A)
and $\beta=0$ (Model B), yielding $1.58^{+0.08}_{-0.09}$ and
$1.66^{+0.07}_{-0.08}$, respectively.
Thus, the model prediction for the slope of the total density profile 
is systematically flatter than isothermal.

%%%%%%%%%% Figure 9
\begin{figure}
\begin{center}
\begin{minipage}{8.4cm}
\epsfxsize=8.4cm 
\epsffile{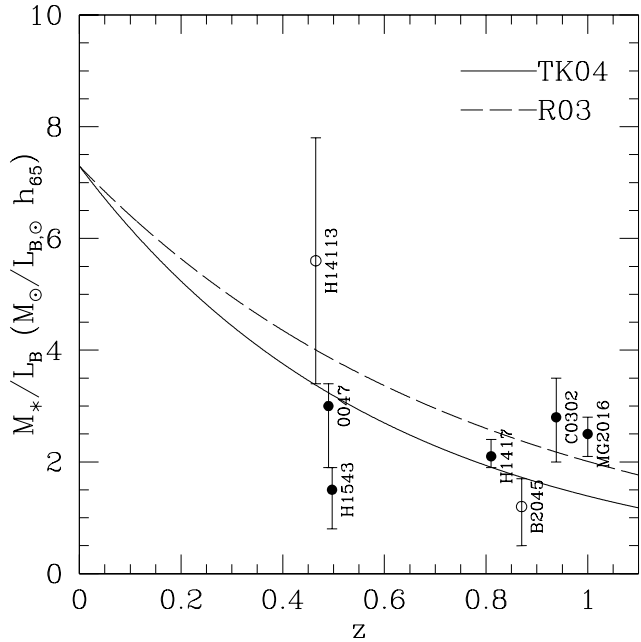}
\end{minipage}
\end{center}
\caption{
Redshift evolution of the stellar mass-to-light ratio.
The open circles correspond to HST~14113$+$5211 and B~2045$+$265
derived from the current lensing model (for an isotropic velocity)
without the FP constraints,
whereas the solid circles are taken from TK04 with the same models.
The solid and dashed lines show
the average evolution of $M_\ast/L_B$ derived from the FP of early-type
lensing galaxies by TK04 and Rusin et al. (2003), respectively.}
\label{fig:ml}
\end{figure}
%%%%%%%%%%

%%%%%%%%%%%%%%%%%% Sec.6 %%%
\section{DISCUSSION AND CONCLUDING REMARKS}

As explored in this work and also in TK04, the mass-to-light ratio for both
of our targets based on the joint lensing and dynamical
analysis is virtually consistent with that obtained from the FP constraint;
the mass-to-light ratios at $z_L = 0.464$ and 0.868 are systematically smaller
than the current average value of
$(M_\ast/L_B)_0 = 7.3 \pm 2.1 h_{65}$~$M_\odot/L_{B,\odot}$
(Gerhard et al. 2001), implying the aging of stellar populations from
these redshifts to the present day. For comparison with other sample lenses
for which the similar analysis has been employed, we plot, in
Figure \ref{fig:ml}, the redshift evolution of the stellar mass-to-light ratio
for both of our targets (open circles) and the TK04 sample
(solid circles).  It is found that the current
sample lenses follow the general redshift evolution of $M_\ast/L_B$, as
guided by the solid and dashed lines showing the average evolution of
$M_\ast/L_B$ derived from the FP of early-type lensing galaxies by TK04
and Rusin et al. (2003), respectively.

The inner slope of the dark-matter halos, $\gamma$, ranges from 0.0 to 1.8.
More specifically, the most likely values of $\gamma$ (with the FP constraints)
for HST~14113$+$5211 are
$0<\gamma<1.6$ ($0<\gamma<1.8$) at $1~\sigma$
for Model A (Model B), whereas for B~2045$+$265, we obtain
$0<\gamma<0.5$ ($0<\gamma<0.8$) at $1~\sigma$ for Model A (Model B).
These values are consistent with the density slopes of dark halos predicted by
cosmological N-body simulations, showing $\gamma$ of 1.0 to 1.5
(e.g., Navarro, Frenk, and White 1995; Moore et al. 1998; 
Fukushige \& Makino 2003 and references therein), although the present-day
inner slope of the dark-matter halos is affected by the effect of
baryonic infall to some extent; the effect seems to be not so significant
as obtained from other lens samples (TK04). It is also drawn that the inner
slope of the dark-matter halos is systematically flatter than
an isothermal profile.

The index for a single power-law model ($\gamma'$) is almost 2 for
HST~14113$+$5211, thereby suggesting that a singular isothermal model
is a good fit to this lensing galaxy. For B~2045$+$265, $\gamma'$ ranges
from 1.5 to 1.7, which is systematically flatter than isothermal
(Figure \ref{fig:B2045_likeli2}).
This flat slope is related to the very small velocity
dispersion of stars $\sigma = 213$ km~s$^{-1}$ compared with
$\sigma_{\rm SIE} = 397$ km~s$^{-1}$; a flatter slope of a dark halo
than SIE is needed to reduce the radial gravitational force and thus
the velocity dispersion of stars. Alternatively, some part of $M(<R_E)$
for B~2045$+$265 is provided by the group of galaxies, thereby causing
a large value of $\sigma_{\rm SIE}$. This effect may partly contribute
to a non-negligible scatter of $\gamma'$ from an isothermal index (2)
as is also reported in other lens samples (TK04).

%%%%%%%%%%%%%%%%%%%%%%%%%%%%%%%%%%%%%%%%%%%%%%%%%%%%%%%%%
\section*{Acknowledgments}
This work has been supported in part by a Grant-in-Aid for
Scientific Research (15540241, 16740118 and 177401166827) 
of the Ministry of Education, Culture,
Sports, Science and Technology in Japan.

%%%%%%%%%% References %%%%%%%%%%%%%%%%%%%%%%%%%%%%%%%%%%

%%%%%%%%%%%%%%%%%%%%%%%%%%%%%%%%%%%%%%%%%%%%%%%%%%%%%%%%%
\appendix
\section{Estimation of the Lens Luminosity in the B band}

For the lensing galaxy of HST~14113$+$5211,
F98 reported the F702W(AB) magnitude as $20.78 \pm 0.05$ mag.
Based on the K-correction using Coleman, Wu, \& Weedman (1980) and
the assumption of no galaxy evolution, F98 derived the absolute B magnitude as
$M_B(AB) = -19.32 + 5 \log h$ for $(\Omega_0,\lambda_0)=(1,0)$ and
$z_L=0.46$.
Then, we transform this magnitude into Vega-based B magnitude
[$M_B({\rm Vega}) = M_B(AB) + 0.12$, Schmidt, Schneider, \& Gunn 1995]
and consider the revised lens redshift of $z_L=0.464$ (Lubin et al. 2000) and
cosmological parameters of $(\Omega_0,\lambda_0)=(0.3,0.7)$
(Spergel et al. 2003). We thus obtain $M_B = -19.60 + 5\log h$, giving
the luminosity of $L_B/L_{B,\odot} = 2.6 \times 10^{10}$
($h=0.65$) for the lensing galaxy of HST~14113$+$5211.

For the lensing galaxy of B~2045$+$265,
Fassnacht et al. (1999) reported the various infrared magnitudes
in a $1.\arcsec 9$ diameter aperture (corresponding to the size of the Einstein
ring in their lens model). Based on the K-correction for an Sa galaxy
as the lens appears to show its typical spectrum,
they arrived at the rest-frame $B$-band luminosity of
$2.36 \times 10^{10} h^{-2} L_\odot$ in this aperture
for $(\Omega_0,\lambda_0)=(1,0)$. To derive the total $B$-band luminosity,
we adopt the work of Rusin et al. (2003), where they obtained the
intermediate-axis effective radius ($R_e$) determined by fitting
the observed brightness distribution to a de Vaucouleurs profile.
Using their $R_e$ of $0.\arcsec 38$ and $(\Omega_0,\lambda_0)=(0.3,0.7)$,
we obtain the luminosity of $L_B/L_{B,\odot} = 1.3 \times 10^{11}$
($h=0.65$) for the lensing galaxy of B~2045$+$265.

\label{lastpage}
\end{document}